# Title: High-Resolution Tunneling Spectroscopy of Fractional Quantum Hall States

**Authors:** Yuwen Hu[1†], Yen-Chen Tsui[1†], Minhao He[1†], Umut Kamber[1], Taige Wang[2], Amir S. Mohammadi[1], Kenji Watanabe[3], Takashi Taniguchi[4], Zlatko Papic[5], Michael P. Zaletel[2], Ali Yazdani[1*]

**Affiliations:**

[1] *Joseph Henry Laboratories and Department of Physics, Princeton University, Princeton, NJ 08544, USA*

[2] *Department of Physics, University of California at Berkeley, Berkeley, CA 94720, USA*

[3] *Research Center for Functional Materials, National Institute for Materials Science, 1-1 Namiki, Tsukuba 305-0044, Japan*

[4] *International Center for Materials Nanoarchitectonics, National Institute for Materials Science, 1-1 Namiki, Tsukuba 305-0044, Japan*

[5] *School of Physics and Astronomy, University of Leeds, Leeds LS2 9JT, UK*

*Correspondence to: yazdani@princeton.edu

† These authors contributed equally to this work

**Strong interaction between electrons in two-dimensional systems in the presence of a high magnetic field gives rise to fractional quantum Hall states that host quasiparticles with fractional charge and fractional exchange statistics. Here, we demonstrate high-resolution**

**scanning tunneling microscopy and spectroscopy of fractional quantum Hall states in ultra-clean Bernal-stacked bilayer graphene devices. Spectroscopy measurements show sharp excitations that have been predicted to emerge when electrons fractionalize into bound states of quasiparticles. We find energy gaps for candidate non-abelian fractional states that are larger by a factor of five than other related systems – for example semiconductor heterostructures – and this suggests bilayer graphene is an ideal platform for the manipulation of these quasiparticles and for the creation of a topological quantum bit. We also find previously unobserved fractional states in our very clean graphene samples.**

**Main Text:**

In two-dimensional (2D) systems, quenching electrons' kinetic energy with the application of large perpendicular magnetic fields enhances the interactions among electrons and drives the formation of fractional quantum Hall (FQH) states (*1*, *2*). FQH states and their low-energy excitations with anyonic statistics have been of interest for many decades (*3–6*); but there is renewed interest in such states because of their potential application for creating topologically protected qubits (*7–9*). Much of what is known about these states and their excitations has been obtained through macroscopically averaged experiments or studies of their edge states (*1*, *2*, *10–16*). Local spatially resolved experiments have the potential to probe exotic quasi-particles, determine sample quality, and achieve the ideal setting in which non-abelian anyons are long-lived for studies of their braiding statistics and for creating topological qubits. Local single electron transistor experiments have already been used to detect fractional charges (*17*, *18*), including *e*/4 charged quasi-particles in the even-denominator FQH states, which are the most promising candidate for hosting non-abelian anyons (*7*). New approaches for local detection of anyons and even testing their fusion rules have been proposed (*19*, *20*). Building on the recent success of atomic scale scanning tunneling microscopy and spectroscopy (STM/S) experiments to study quantum Hall states in monolayer graphene (*21*, *22*), here we show the application of these techniques to the myriad of FQH states in Bernal-stacked bilayer graphene (BLG). In this work, STM's ability to identify ultra-clean devices is harnessed to discover remarkably large energy gaps protecting FQH states, including those that are consistent with the proposed Moore-Read paired state hosting non-abelian anyons (*7*). Our tunneling spectroscopy experiments probe higher energy FQH's excitations, which are beyond the reach of transport studies, to reveal sharp peaks over narrow energies that have been predicted theoretically (*23*). Surprisingly, even though FQH states are highly correlated phases in which their low-energy excitations are distinct from electrons, high-energy composite states of fractionalized excitations have a strong overlap with electron or hole-like excitations. The ability to perform atomic scale imaging and spectroscopy allows us to probe the influence of single atomic defects and to characterize disorder-free intrinsic bulk properties of FQH states for the first time, which shows remarkably good agreement with theoretical calculations. The exponential suppression of thermally activated quasi-particles due to the large energy gaps reported here for the non-abelian states in BLG makes BLG an ideal platform for the realization of topological qubits based on non-abelian anyons.

The observation of FQH states is ubiquitous to materials platforms with 2D electronic states (*1*, *2*, *10–14*). However, their presence in graphene-based systems at the surface of devices, that can be made pristine on the atomic scale, makes high-resolution STM studies of FQH states possible (*21*, *24*). The Coulomb interaction in graphene-based systems in a magnetic field is also relatively stronger than, for example, that in GaAs because of the dielectric environment for these devices, making the energy scales for the formation of FQH states in graphene-based systems larger in comparison. Previous transport and electronic compressibility experiments have established that at partial filling of the lowest Landau levels (lowest LL) of BLG, in which the accidental near-degeneracy between orbital quantum number $N = 0$ and $N = 1$ LLs is lifted, this material system shows both odd (in $N = 0$) and even denominator FQH states (in $N = 1$) (*11, 12, 25–29*). Depending on experimental conditions, in particular the value of the displacement field perpendicular to the bilayer, this system can realize different broken symmetry states (*30, 31*), from which various FQH states emerge (*12, 25–29*). Although information about the broken symmetry states has been extracted from various macroscopically averaged experiments (*30, 31*), visualization of such broken symmetry states, as recently achieved in monolayer graphene (*21, 22*), has not been performed.

For our experiments, we fabricated BLG devices using hexagonal boron nitride (hBN) substrates with graphite back gates and performed various cleaning procedures to create devices with atomically pristine areas as large as 200 nm × 200 nm, without any defects or adsorbates (see Methods for sample fabrication details). The STM topography of such ultra-clean devices shows the moiré pattern due to a mismatch of BLG and hBN lattices (6.7 nm moiré periodicity) and very weak longer-range variations, see Fig. 1a and Fig. 1b for a device schematic and STM topography. Since the BLG in our devices is in an asymmetric dielectric environment (hBN on one side), we observe an energy splitting of $U = 30$ meV between the states occupying different layers (Fig. 1c). We attribute this to an effective intrinsic displacement field $D_0$, which induces a gap in the local density of states (LDOS) between the conduction and the valance band of BLG at zero magnetic field. As shown in Fig. 1c, the zero-field differential conductance d$I$/d$V$ as a function of sample bias $V_B$ measurements shows such a gap in the LDOS, which is independent of the gate voltage ($V_G$), except when we approach charge neutrality near $V_G = 0$, where it is strongly enhanced likely due to exchange effects (see Fig. S1 in SI for larger gate range). A single particle calculation of the LDOS of the top layer BLG using a tight-binding model with an interlayer potential of $U = 30$ mV, due to $D_0$, can capture the gap and LDOS we observe at zero field away from charge neutrality (see SI discussion I, Fig. S2-3). This comparison confirms that STM probes the electronic properties of BLG through measurements of the layer closest to the STM tip (layer separation is $d = 0.35$ nm). The corresponding $D_0 = \frac{\epsilon U}{d} = 343$ mV/nm, if we assume a dielectric constant $\epsilon = 4$, is consistent with previous reports of similar device geometry (*12*, *32*), from which the $D_0$ (with $U = 30$ mV) is estimated to be about 300 mV/nm (*33*). Both numbers are much larger than the displacement field generated by back gate $V_G$ in our studies. As we describe below, the presence of large intrinsic $D_0$ results in layer polarization of the FQH states and influences our STM measurements.

Application of a perpendicular magnetic field reconstructs the LDOS of BLG into discrete LLs, which appear as lines in gate-tunable scanning tunneling spectroscopy maps with 8 mV widths. With increasing $V_G$, after the LL closest to the Fermi energy ($E_F$) is filled, the higher LLs transit from unoccupied ($V_B > 0$) to occupied state ($V_B < 0$) (Fig. 1d at 5 T, see Fig. S4 in SI for other fields). These transitions mark when each of the four-fold degenerate LLs, which we denote as filling factor $\nu = -8, -4, 4, 8$ relative to charge neutrality, are fully occupied. Like previous work on monolayer graphene (*21*), tip preparation and sample cleanliness result in a minimal tip-gating effect (see Methods), as the LLs appear parallel to the $V_G$ axis. These measurements show gapped integer quantum Hall (IQH) states as well as broken symmetry quantum Hall ferromagnetic (QHFM) phases when the four-fold degeneracy of the IQH states is lifted (*34*, *35*). From these features, the filling factor $\nu$ and the orbital number $N$ can be determined (as labeled in Fig. 1d). Further analysis (right panel in Fig. 1d, also see discussion II, and Fig. S5 in SI) shows the LLs energies to follow the expected $E_N = \hbar\omega_c \sqrt{N(N-1)}$ sequence with extracted cyclotron gap $\hbar\omega_c$ at different fields (*36*, *37*), which also identifies different orbitals and yields an effective mass of $m^* \approx 0.044 m_e$, both consistent with previous studies (*36*, *38*). Details of the field dependence of various gaps are discussed in SI (see Fig. S6).

We focus on the spectroscopic properties of the lowest LLs ($N = 0, 1$) by first examining signatures of layer polarization in the filling range $-4 < \nu < 4$, which has been explored in macroscopic measurements before (*31*) but can be visualized in our STM measurements. These experiments also determine the orbital states that are being filled which dictates the sequence of FQH states that can be formed at different fillings in BLG. As spectroscopic measurements show (at 10 T, Fig. 2a), for fillings $-4 < \nu < 0$ the LDOS near $E_F$ of the top graphene layer, which is what STM probes in our experiments, is strongly suppressed relative to other energies. There are multiple features/splitting in the spectra at higher energies for this range of filling, which, as previous work on monolayer graphene demonstrates, requires further investigation to identify their nature (*39*). The suppression of the top layer LDOS near $E_F$ for $\nu < 0$ subsides once we dope the sample to $\nu > 0$. This behavior suggests that in the presence of intrinsic $D_0$, from the manifold of possible $N = 0, 1$ states, those that reside in the lower graphene layer of the BLG (Fig. 2b) are filled first for $\nu < 0$, thereby explaining why the LDOS near $E_F$ measured on the top layer is suppressed in this regime. Once these levels of the bottom layer are filled, the $N = 0$, 1 states of the top layer, which easily couple to the STM tip, contribute to the LDOS near $E_F$.

Another measurement that confirms this picture and provides further information on the sequence of orbital states being filled as a function of density is that of spectroscopic mapping of LLs in the top layer. As shown in Fig. 2b, the single-particle wave function of different orbital state $N$ in each layer, $|\pm N\sigma\rangle$ ($\pm$ refers to K and K' valleys, and $\sigma$ refers to spin) are expected to have different amplitude on the four atomic sites of BLG (A, B, A' and B', as labeled in Fig. 2b) (*36*). While the state $|-0\sigma\rangle$ has no amplitude on the top layer, $|-1\sigma\rangle$ has some spectral weight on the B site of the top layer. We also note that the spectral weight from both LL orbitals in the top layer $|+N\sigma\rangle$ reside in the A sublattice of the top layer. We experimentally find that the relative weak spectral features near $E_F$ for $\nu < 0$ (dashed orange box in Fig. 2a, which gets clearer when we bring the tip closer to the sample) reside on B sublattice in spectroscopic maps (Fig. 2c left). In contrast, spectral features near $E_F$ for $\nu > 0$, all reside on the A sublattice (Fig. 2c right), which is consistent with the layer polarization picture. In addition, for $\nu < 0$, near the Fermi energy (which is interrupted by a Coulomb gap), we see B-sublattice polarization in $-3 < \nu < -2$ and $-1 < \nu < 0$, suggesting that $|-1\sigma\rangle$ is being filled in these filling ranges. This finding suggests sequential $N = 0, 1, 0, 1$ filling for the bottom layer for $\nu < 0$. Measurements of the FQH states described below, further confirms this picture and show that the same sequence of filling is followed for $\nu > 0$. Our results are consistent with the high displacement field limit of a previous study (*31*).

Signatures of FQH states appear in high-resolution STM spectroscopy of LDOS near $E_F$, where experimental measurements are influenced by both the Coulomb gap ($\Delta_C$) for the addition/removal of an electron to/from 2D systems at a high magnetic field (*21, 40–42*) and the energy to create quasi-particles in various FQH states. At generic fillings, the system is expected to be compressible; nevertheless, a Coulomb gap arises in tunneling because the low-energy excitations are orthogonal to the injection of a single electron (*43*). Figure 3a shows such measurements at 13.95 T for $0 < \nu < 1$, where we expect to fill the $|+0\sigma\rangle$ state, we find spectroscopic features at filling factors $\nu = \frac{p}{2p+1}$ ($p \in \mathbb{Z}$, with $p$ up to 8 and -9) corresponding to the Jain sequence (two-flux composite fermions states) (44) that appear together with broad suppression near $E_F$ corresponding to $\Delta_C$. We find a similar sequence of FQH states features for -4 < ν < -3, -2 < ν < -1, and 2 < ν < 3 (See discussion III, Fig. S7-8 in SI), where FQH states are observed at $\nu_{eff} \equiv \nu - \lfloor \nu \rfloor = \frac{p}{2p+1}$ at consecutive integer p, consistent with the expectation that $|\pm 0\sigma\rangle$ states are partially occupied at these fillings.

The onset of FQH states is accompanied by several distinct features in the high-resolution gate-tunable scanning tunneling spectroscopy measurements. First, we see that, in contrast with the tunneling spectra at partial fillings away from FQH states which show broad gap-like features associated with $\Delta_C$, the FQH states are marked by enhanced threshold tunneling gaps $\Delta_t$, which is flanked by sharp peaks in the spectra (see Fig. 3b, for example). Second, the additionally required charge excitation energy to add an electron (hole) can be evaluated by the sharp energy jumps $\Delta_e$ ($\Delta_h$) upon entering (exiting) the incompressible FQH gaps. In a perfectly clean system, the gate range over which the enhanced tunneling gap $\Delta_t$ is observed is expected to be equal to the chemical potential jump $\delta\mu$ (e.g., the thermodynamic gap) required to tune the system across these incompressible states' gaps (39), because in an incompressible state, $\mu$ decreases in exact proportion to $eV_G$. However, in the presence of impurities, the gate range can overestimate the chemical potential jump $\delta\mu$ because the filling of in-gap impurity states partly screens the applied gate voltage. As we describe below, our tunneling spectroscopy can in fact be used to determine the role of in-gap states in our sample, and place a lower bound on the thermodynamic gap in a realistic setting. The schematic shown in the inset of Fig. 3d depicts the key features of the spectra $\Delta_e$, $\Delta_h$, $\Delta_t$, $\Delta_C$, $\delta V_{G,h}$, $\delta V_{G,e}$ (there are slight differences between gate ranges for electron addition or removal), as well as the absolute value of the slope of spectroscopic features $S_{G,h}$, $S_{G,e}$ within the region of enhanced tunneling gap.

While a jump in the chemical potential with the appearance of incompressible FQH state is not surprising, the observation that FQH states, which do not have electron-like low energy quasi-particles, show sharp resonances in electron tunneling spectra is remarkable. Previous theoretical studies had anticipated that such sharp features in tunneling experiments may occur as high angular momentum excitations of fractional quasi-particles of FQH states (*23*); however, they have not been detected in any previous tunneling spectroscopy experiments. Conceptually, the added tunneling electron (hole) can be thought to be transformed into composite states of $2p + 1$ fractional quasi-particles (quasi-holes) and the Coulomb repulsion among which is compensated by them being in a high angular momentum state (*23, 40, 43*). We discuss the experiment for testing the validity of the FQH state features in the tunneling spectra in Methods.

To quantitatively analyze our spectroscopic measurements FQH states, we extract from our gate-tunable scanning tunneling spectroscopy measurements the values of $\Delta_e, \Delta_h, \Delta_t, \Delta_C$, $\delta V_{G,h}, \delta V_{G,e}, S_{G,h}$ and $S_{G,e}$, (at $B = 13.95$ T) for all the odd denominator FQH states in the range $0 < \nu < 1$ (See Method for detailed definitions and extraction procedures). Examining measurements at different fields (see Fig. S11 in SI), we find that $\Delta_C$ follows the expected $\sqrt{B}$ behavior of the Coulomb energy (*40*), while the $\Delta_t$ of the various FQH states systematically increases in its magnitude beyond $\Delta_C$ in a linear fashion with $B$ field (see Fig. 3c). Motivated by clear enhancement beyond $\Delta_C$ for tunneling into the FQH states, we plot $\Delta_e + \Delta_h$ as a function of $p$ (where $p = \frac{\nu}{1-2\nu}$) and find its systematic suppression with increasing $|p|$ (Fig. 3d). An interesting experimental finding is that $\Delta_e + \Delta_h$ follows a trend of $\frac{1}{2p+1}$ as a function of $p$ (Fig. 3d green solid lines) for states with $p > 0$, while for $p < 0$ these gaps do not show the same functional form but still monotonically decrease for larger $|p|$. This $p$ dependence resembles the previously predicted trend for the gap to create a quasi-particle-quasi-hole pair for the odd denominator FQH states within the composite fermions theory (*45*, *46*). However, it is not a priori obvious there should be such a resemblance, as $\Delta_{e/h}$ represents a different quantity: the increased energy required to inject/remove an entire electron upon entering the FQH gap. This correspondence thus requires further investigation. We also note that $\Delta_e + \Delta_h$ has an asymmetry between positive and negative $p$, and the $\nu = 2/3$ state ($p = -2$) shows an anomalously large gap compared to other FQH states. This reflects the particle-hole symmetry breaking in the zeroth Landau level of bilayer graphene and is consistent with the result by other probes (*12*, *28*). The observations of a combination of sharp features in spectroscopy and the hierarchy of $\Delta_e + \Delta_h$ among the odd denominator FQH states demonstrate a new spectroscopic approach for studying these states by characterizing the process of fractionalizing the tunneling electron (hole) into quasi-particles (hole) of FQH states.

The local measurements of $\delta V_{G,\text{avg}} = (\delta V_{G,e} + \delta V_{G,h})/2$, which shows a similar trend as a function of $|p|$ (See Fig. S12), can be analyzed to characterize $\delta\mu/e$ and the thermodynamic gaps for various FQH states even in the realistic setting where there is a background of impurities in the sample. In the absence of any impurities, $e\delta V_{G,\text{avg}} \approx \delta\mu$, and changing the gate voltage would shift the chemical potential of the sample in such a way that all features in the incompressible state shift linearly with the absolute value of the slope $S_{G,e} = S_{G,h} = 1$ as a function of the gate voltage. However, in the presence of any impurities that can electrostatically influence the areas under the tip (which can be one in a few hundred nm square areas, see SI section V and Methods), sweeping the gate voltage fills these states and in turn changes the slope $S_G$ to be less than one. Using the combination of measurements to determine the product $eS_G\delta V_G = e(S_{G,e}\delta V_{G,e} + S_{G,h}\delta V_{G,h})/2$ puts a lower bound on the thermodynamic gaps $\delta\mu$ for various FQH states in the realistic setting of the sample with a finite concentration of defects (Fig. 3d and SI section V). The close correspondence between $(\Delta_e + \Delta_h)/2$ and $eS_G\delta V_G$ apparent from Fig. 3d is equivalent to the fact that the tunneling threshold is more or less constant on either side of the FQH state. We emphasize that this did not, a priori, need to be the case, and future work to understand this correspondence is in order.

Extending our measurements to when $|\pm 1\sigma\rangle$ states are being filled, we find spectroscopic signatures of the exotic even-denominator FQH states, a different hierarchy of quasi-particle excitation gaps, as well as some unexpected FQH states. As shown in Fig. 4a, in the range $-1 < v < 0$ when filling $|-1\sigma\rangle$ state, a pronounced even denominator state at $v = -1/2$ (or $v_{\text{eff}} \equiv v - \lfloor v \rfloor = 1/2$, defined for the simplicity of the discussion) in the spectroscopy measurements is flanked by other FQH states (inset of Fig. 4a). The extracted $eS_G\delta V_G$ corresponding to the lower bound of the thermodynamic gaps for this sequence of FQH states shown in Fig. 4b show several remarkable features. First, the appearance of $v_{\text{eff}} = 1/2$ flanked by $v_{\text{eff}} = 8/17$ and $7/13$ strongly suggests that the even denominator state corresponds to the non-abelian Moore-Read Pfaffian state (7), together with its Levin-Halperin daughter states forming on its sides (47). This observation (and lack of any features at $v_{\text{eff}} = 9/17$ and $6/13$) is also consistent with transport and compressibility studies (28, 29) favoring Pfaffian, over anti-Pfaffian (48) or PH-Pfaffian (49), in BLG. However, the second important feature of the data in Fig. 4b is that both the $(\Delta_e + \Delta_h)/2$, as well as the lower bound on the thermodynamic gap $eS_G\delta V_G$ for the measured $v_{\text{eff}} = 1/2$ locally (Fig. 4c) are a factor of 5 larger than those from previously reported macroscopically averaged experiments (26, 28). Other remarkable features of the data in Fig. 4a are the observation of $v_{\text{eff}} = 2/5$ and $3/5$ states that may be Jain sequence due to a small mixture of $N = 0$ state or the predicted Read-Rezayi states (50), which like the

Moore-Read state are also expected to be non-abelian (*51*). There is also the observation of new and unexpected FQH states at $\nu_{eff}$ = 5/11, 6/11, and 5/9. Experiments at fillings $-3 < \nu < -2$, $1 < \nu < 2$, and $3 < \nu < 4$ all show clear signatures of even-denominator states, with the latter two ranges also showing clear signatures of daughter states consistent with the even-denominator states being Pfaffian (see section IV, and Fig. S13-15 in SI).

A natural question is how the properties of FQH states vary spatially and what are the typical spatial variation of our spectroscopic measurements in our BLG samples. In general, we find that the moiré structure due to h-BN underneath does not affect our measurement. However, in some regions of the sample, there are long-range variations of the spectra, which in some cases can be clearly attributed to the dilute concentration of defects in our sample. For example, Fig. 4d, 4e show that a sub-surface defect locally suppresses the even-denominator state at $\nu$ = 3/2 ($\nu_{eff}$ = 1/2) with the $\Delta_t$ recovering from on the scale of 100 nm away (Fig. S16 in SI). The local suppression of the incompressible FQH gap may arise from the trapping of quasi-particles or quasi-holes near the defect, but remains to be fully understood. Future studies can use spatial resolved spectroscopy to further investigate the distribution of such variations and the impact of the proximity of various impurities in the sample to each other on the spectroscopic properties of FQH states. The concentration of defects observed in our field of view also provides a consistent picture of how these states influence the gate range for being in the incompressible gap as described above (See SI section V).

The remarkably large energy scales for quasi-particle excitations and lower bounds on thermodynamic gaps we have extracted from our measurements away from any defects in our ultra-clean samples suggest that we are probing FQH states with unprecedented precision, thereby motivating comparison of our results with those of idealized theoretical calculations.

Numerical simulation of the LDOS using screened Coulomb interactions that are appropriate for our one-sided-hBN and a single gate can be performed within the standard framework for describing FQH states (see section VI in SI). The exact diagonalization (ED) calculations show signatures of the sharp features similar to our tunneling experiments (Fig. S17 in SI), and the tunneling threshold energy on different system sizes that can be used to determine $\Delta_t$ in the limit of large system size (Fig. S20 in SI). Given that the theory ignores the particle-hole asymmetry, we compare the theoretical results with the average value of the experimentally measured $\Delta_t$ for two equivalent FQH states relative to half-filling. In Fig. 4f, we show that using a reasonable value for the dielectric constant $\epsilon$, we find measurements of $\Delta_t$ for FQH states in both $N = 0$ and 1 states are in excellent agreement with theory (see section VIII, Fig. S8c, S13d, S14d, S15d, Table S10 in SI for comparison at other fillings and see section VIII, Table S9 in SI for small changes in $\epsilon$ as a function of LL fillings). Similarly, we can compare our measurements of $eS_G \, \delta V_G$ for some of the FQH states to ED and the density matrix renormalization group calculation of the thermodynamic gaps and find them to somewhat smaller than the idealized situation as expected (see section VI, VII, and VIII, Table S11 in SI). However, the remarkably large lower bound of thermodynamic gaps for exotic FQH states, such as close to 19 K at 13.95 T for the Moore-Read states in our local measurements of ultra-clean BLG devices show that this is an ideal system for exploring the properties of the predicted non-abelian anyons. In a perfectly clean system, we expect the relation of the thermodynamic gap $\Delta_\mu$ and quasi-particle gap $\Delta_{qp}$ following $\Delta_{qp}=\Delta_\mu/4$ for the Moore-Read even-denominator state. Our result then corresponds to quasi-particle gap $\Delta_{qp}$ about 4.7 K, which is comparable to the transport activation gap from a coincident experiment (*52*). In ultra-clean GaAs devices (*53*), the gaps for such states realized at $\nu = 5/2$ are about a factor of 5 smaller due to its appearance limited to lower magnetic fields and increased dielectric screening. Future STM experiments in ultra-clean BLG samples can utilize electrostatic confining potentials to directly visualize the fractionalization process which has unique signatures for each FQH state (*20*).

STM studies also provide a more precise way to characterize FQH states, as well as an opportunity to find previously undiscovered states. In GaAs, estimates of the bulk gap for the FQH state from transport studies have been consistently smaller than those predicted theoretically (*54*), likely because such measurements are sensitive to disorder in the sample (*55*) (see above for an example of the local influence of individual impurities on FQH gaps, and Fig. S16 in SI, as well as discussion on the influence of in-gap states on estimating the lower bound on thermodynamic gaps). A coincident experiment (*52*) reports the thermodynamic gaps for the FQH states examined here in ultra-clean devices of BLG, where they find comparable gaps using a graphene sensor, to our lower bound of thermodynamic gap for the Jain sequence reported here (see SI Fig. S27). Our measurements of $\nu_{eff} = 1/2$ state locally with the STM still find a larger lower bound compared to these measurements, a behavior that is not surprising given that STM can make measurements in locations that are furthest away from defects to find the best experimental conditions closest to the idealized conditions for this state, which could be more sensitive to local disorder.

STM's ability to probe the disorder-free areas of the sample can also uncover new FQH states. We encountered an example of such a finding while probing the higher $N = 2$ LL, where we find not only a rich sequence of odd denominator states (*32*) but also an unexpected even denominator FQH state (Fig. 5a). While the even denominator FQH state in higher LLs has been reported in monolayer graphene (*56*), it is the first time such a state has been observed in a higher-LL of BLG. The hierarchy of local gaps (Fig. 5b) suggests the odd denominator FQH states are the composite fermion states. Interestingly, both $\Delta_e + \Delta_h$ and $eS_G \delta V_G$ are more particle-hole symmetric with respect to half filling than in the lowest LLs, consistent with the previous report (*32*). Different from the stripe phase observed in higher LLs in GaAs (*57, 58*), the new even denominator FQH state may be due to either Moore-Read (*7, 51*), parton 221 state (*56*) or other exotic composite fermion paired state (*59*). The exact isospin flavor, the role of the $N = 2$ LL's spinor structure and a microscopic description of such a state remain interesting open questions. Overall, our studies established that ultra clean BLG is an ideal platform for future local experiments on FQH states and for potentially creating a topological qubit using local manipulation of non-abelian anyons in the even-denominator FQH states.

**Acknowledgments:** This work was primarily supported by DOE-BES grant DE-FG02-07ER46419, the Gordon and Betty Moore Foundation's EPiQS initiative grants GBMF9469 to A.Y. Other support for the experimental work was provided by NSF-MRSEC through the Princeton Center for Complex Materials NSF-DMR- 2011750, NSF-DMR-2312311, ARO MURI (W911NF-21-2-0147), ONR N00014-21-1-2592, and ONR N000142412471. A.Y. acknowledge the hospitality of the Aspen Center for Physics, which is supported by National Science Foundation grant PHY-1607611, where part of this work was carried out. M.Z. and T.W. are supported by the U.S. Department of Energy, Office of Science, Office of Basic Energy Sciences, Materials Sciences and Engineering Division, under Contract No. DE-AC02-05CH11231, within the van der Waals Heterostructures Program (KCWF16). Z.P. acknowledges support by the Leverhulme Trust Research Leadership Award RL-2019-015 and in part by grant NSF PHY-2309135 to the Kavli Institute for Theoretical Physics (KITP). We thank useful conversations with Andrea Young and Amir Yacoby.

# Figure 1

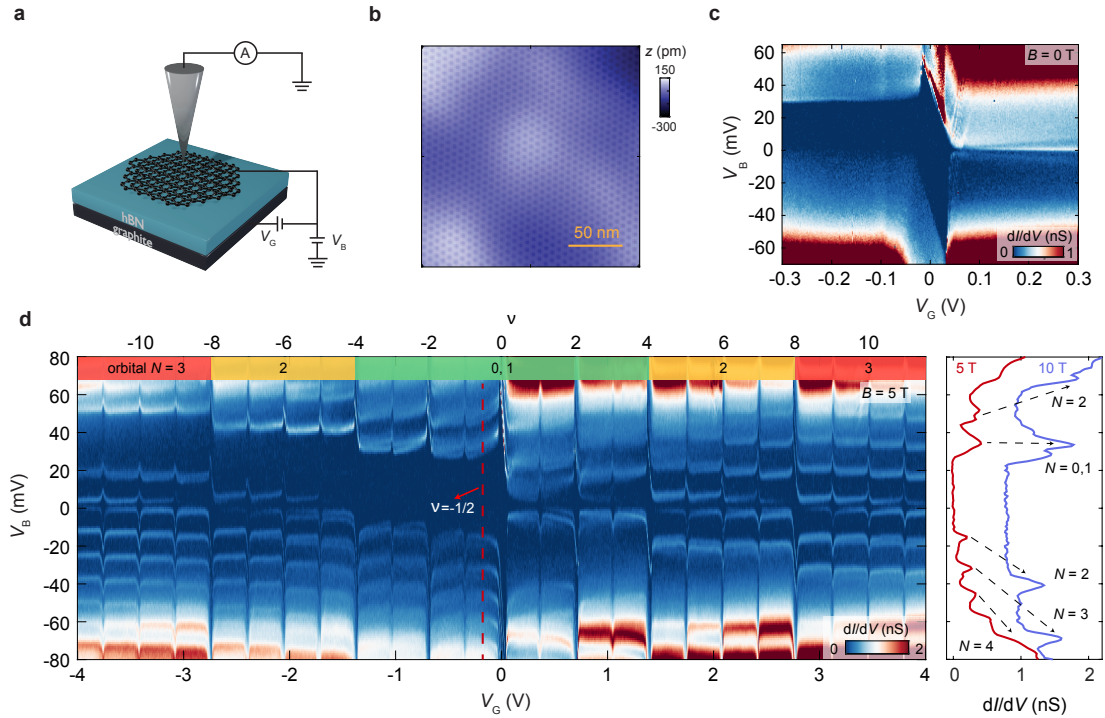

**Figure 1| Experimental setup and tunneling spectra at zero and low magnetic fields**. **a,** The schematic of the bilayer graphene device and the STM/STS experimental setup. A bias voltage $V_B$ is applied to BLG to enable tip-sample tunneling. A voltage $V_G + V_B$ is applied to the bottom graphite gate to electrostatically tune the carrier density. **b,** The topography image of the ultra-clean bilayer graphene device. The scale bar is 50 nm. **c,** The tunneling spectra at zero magnetic field $B = 0$ T. Away from the charge neutrality point at $V_G < 0$ side, a constant gap around 30 mV is observed. Near the charge neutrality point, the gap is enhanced to be 73.5 meV, likely due to exchange interaction. **d,** Left: The tunneling spectra at a low magnetic field $B = 5$ T. The peaks in d$I$/d$V$ spectrum at fixed $V_G$ present narrow LL-like charge excitations. They remain unchanged in energy while increasing the charge density (i.e. parallel to the $V_G$ axis as $V_G$ increases), therefore demonstrating minimal tip gating effect due to the applied bias $V_B$. The LL filling factor $\nu$ and the corresponding orbital numbers are denoted in the top $x$-axis. Right: The d$I$/d$V$ spectrum at filling factor $\nu = -1/2$ at $B = 5$ T (line cut in the left panel indicated by the red dashed line) and $B = 10$ T respectively, with a horizontal offset for clarity. Each peak is labeled by its corresponding orbital number $N$.

# Figure 2

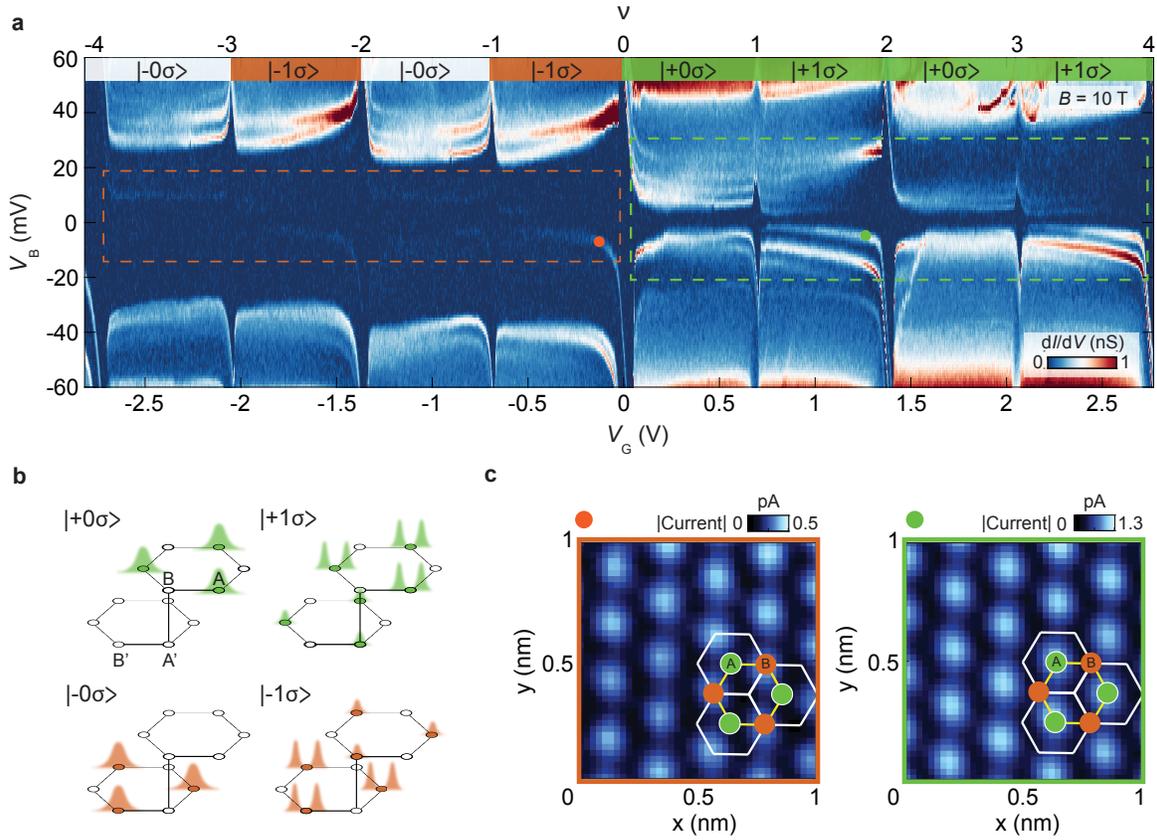

**Figure 2| Atomic wave function imaging of layer polarized $N = 0, 1$ Landau levels at $B = 10$ T. a,** The tunneling spectra of the lowest Landau levels (lowest LLs) at magnetic field $B = 10$ T. The isospin flavors of each LL being filled at the Fermi level are labeled by their corresponding quantum numbers of valley (+, -), orbital number ($N = 0, 1$), and spin ($\sigma$). **b,** The atomic wave function of the lowest LLs with different isospin flavors in BLG. The four atomic sites in the unit cell (A, B, A', B') are labeled. **c,** Atomically-resolved wave function images with sublattice polarization. Left: B sublattice polarized (dimer site on the top layer, in orange) measured at $V_B = -7$ mV, $V_G = -0.128$ V, orange dot in **a**. Right: A sublattice polarized (non-dimer site on the top layer, in green) measured at $V_B = -5$ mV, $V_G = 1.264$ V, green dot in **a**. Note both maps are plotted with absolute current (since $V_B < 0$). Low energy charge excitations in panel **a** are grouped by their corresponding sublattice polarization using the same color scheme (green and orange dashed box).

# Figure 3

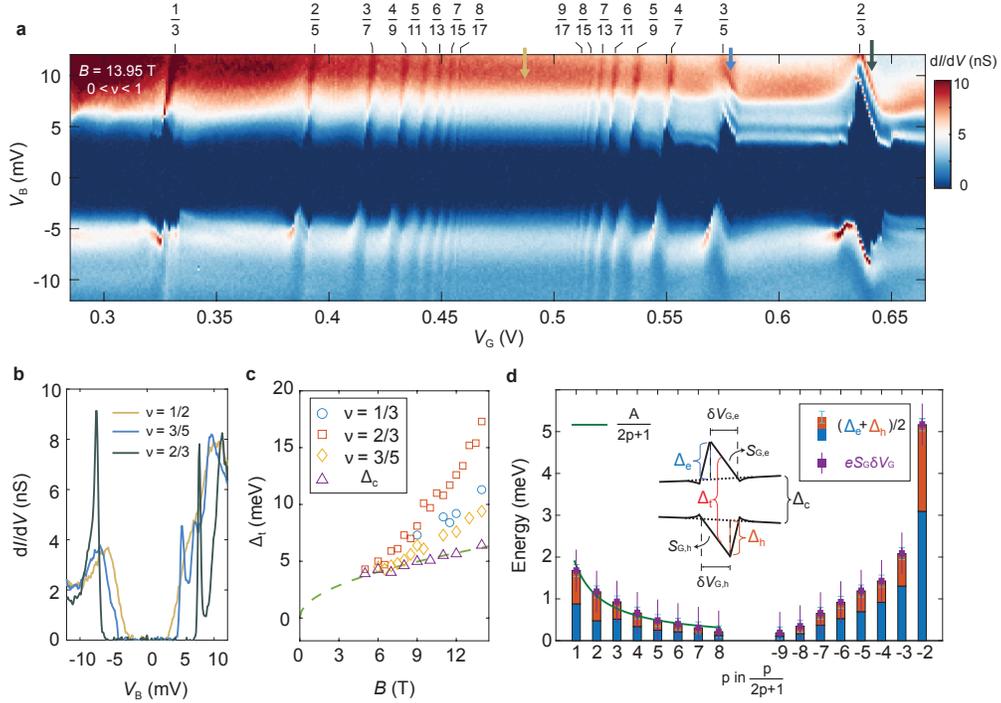

**Figure 3| FQH states within $0 < \nu < 1$ in the $N = 0$ LL. a,** The tunneling spectra of the $N = 0$ LL within $0 < \nu < 1$ at magnetic field $B = 13.95$ T. FQH states following the Jain sequence up to $\nu = 8/17$ and $9/17$ are denoted on the top. **b,** The tunneling spectra at $\nu = 1/2$, and inside the $\nu = 3/5$ and $2/3$ FQH states, indicated by the arrows in panel **a**. **c,** The extracted tunneling gaps $\Delta_t$ of FQH states $\nu = 1/3, 2/3$, and $3/5$ at different magnetic fields $B$. The Coulomb gap $\Delta_C$ at different magnetic fields is also measured and follows the trend of $\Delta_C \propto \sqrt{B}$ (green dashed line). **d,** The FQH states local gaps extracted at filling factor $\nu = \frac{p}{2p+1}$ in the $N = 0$ LL. $(\Delta_e + \Delta_h)/2$ represents the averaged extra excitation energy required to add an electron or a hole to the incompressible FQH states respectively. The blue bars and orange bars represent the portion of the additional energy for an electron and hole respectively. The green solid lines show the fitted $\Delta_e + \Delta_h$ following the trend, $\Delta_e + \Delta_h \propto \frac{1}{2p+1}$ for $p > 0$. The averaged gate width $S_G \delta V_G = (S_{G,e} \delta V_{G,e} + S_{G,h} \delta V_{G,h})/2$ (purple squares) of the incompressible FQH states captures the lower bound of the thermodynamic gap $\delta \mu \cong e S_G \delta V_G$. The inset shows the schematic of the local gaps. Error bars are added to represent the measurement's resolution.

# Figure 4

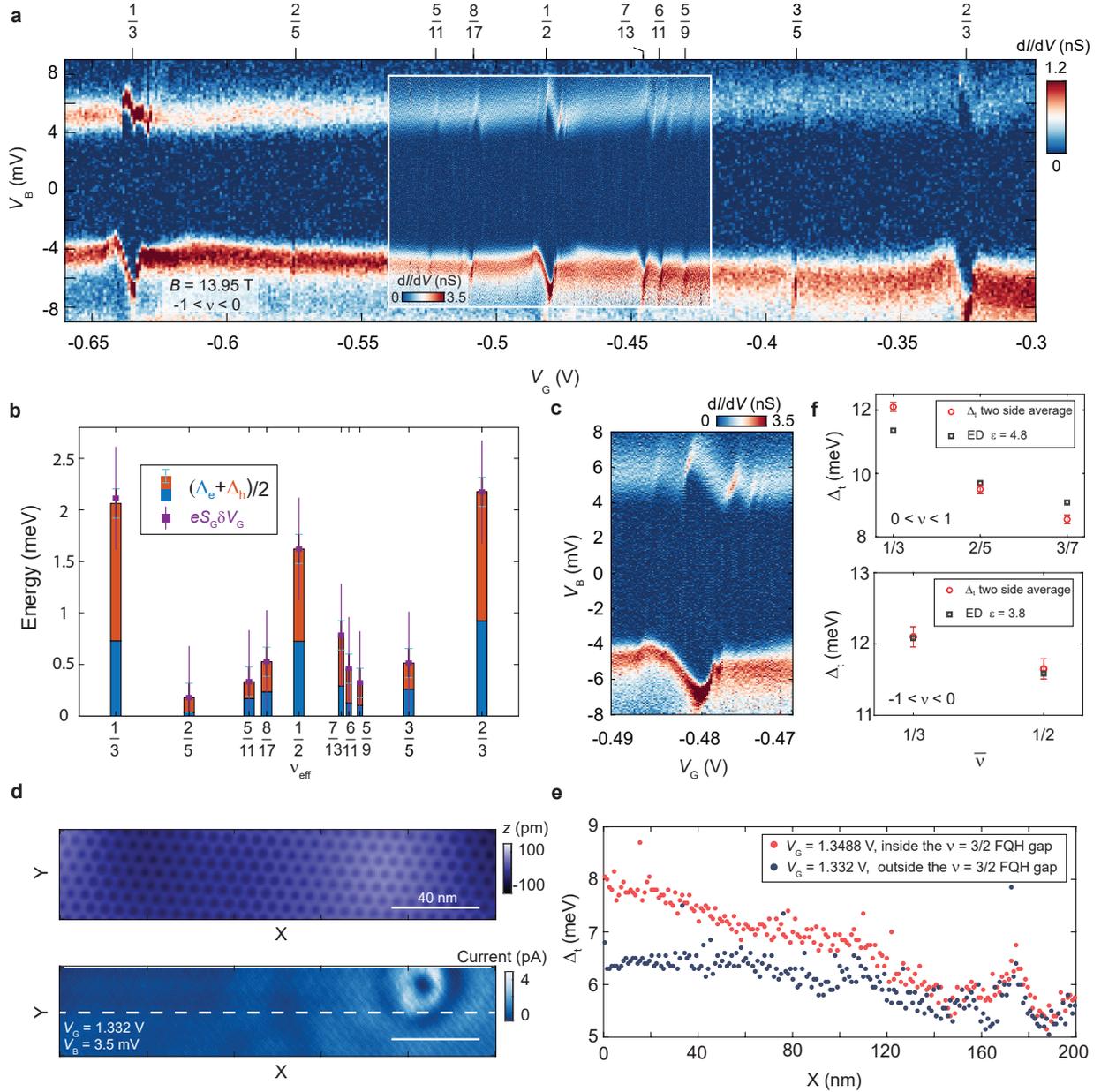

**Figure 4| FQH states in the $N = 1$ LL. a,** The tunneling spectra of the $N = 1$ LL within $-1 < \nu < 0$ at magnetic field $B = 13.95$ T. The corresponding effective fillings factor $\nu_{\text{eff}} = \nu - \lfloor \nu \rfloor$ of each FQH state is denoted on the top. Inset is the zoom-in high-resolution spectra near $\nu_{\text{eff}} = 1/2$. **b,** The local gaps $(\Delta_e + \Delta_h)/2$ (blue and orange bars), and the lower bound of the thermodynamic gap $eS_G\delta V_G = e(S_{G,e}\delta V_{G,e} + S_{G,h}\delta V_{G,h})/2$ (purple squares) of the FQH states observed in panel

**a.** Here, error bars are added to represent the measurements resolution. **c,** The zoom-in tunneling spectra near $v_{eff} = 1/2$ show the even-denominator FQH state with a large thermodynamic gap $eS_G \delta V_G \approx 18.9$ K. **d,** Topography image ($V_B = 0.4$ V, $I = 3$ pA) of a 40 nm by 200 nm clean area. The tunneling current map is taken at setpoint $V_B = 3.5$ mV, $I = 1$ nA following the trajectory recorded at $V_B = 200$ mV, $I = 1$ nA at $V_G = 1.332$ V in the same location as shown in the top panel. The current map reveals disorder with ring-like structures around it. **e,** The spatial variation of the extracted tunneling threshold gap $\Delta_t$ across the linecut (white dashed line in panel D), at two specific gate voltage inside/outside (red/blue) the gap feature of $v = 3/2$ FQH state ($v_{eff} = 1/2$) in another $N = 1$ LL. (The measurement is performed on $v = 3/2$ FQH state due to its signal being stronger on the top layer, thus easier for STM to access.) Both **d** and **f** are measured at $B = 13.2$ T. **f,** Comparison between the measured tunneling gaps $\Delta_t$ (red circles) in the middle of the gap feature and the exact diagonalization (ED) calculations (black squares) for FQH states at $0 < v < 1$ in the $N = 0$ LL (top) and $-1 < v < 0$ in the $N = 1$ LL (bottom). The dielectric constants are determined to be $\epsilon = 4.8$ and $\epsilon = 3.8$ respectively (see SI for details). The tunneling gaps $\Delta_t$ at $\bar{v}$ are averaged result at $v_{eff}$ and $1 - v_{eff}$ (except $v_{eff} = 1/2$) to reach a fair comparison with the ED calculations, which assume particle-hole symmetry. Here, error bars are added to represent the measurement's resolution.

# Figure 5

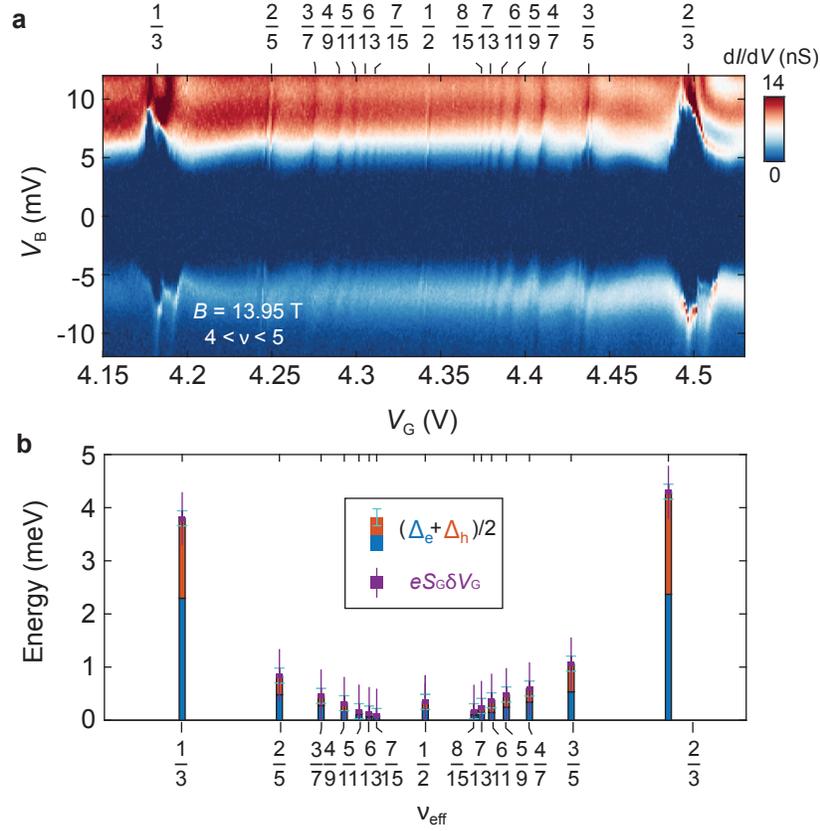

**Figure 5| FQH states within $4 < \nu < 5$ in the $N = 2$ LL. a,** The tunneling spectra of the $N = 2$ LL within $4 < \nu < 5$. A series of odd-denominator FQH states in the sequence of $\nu_{\text{eff}} = \nu - \lfloor \nu \rfloor = \frac{p}{2p+1}$ up to 7/15 and 8/15, and an even-denominator FQH state at $\nu_{\text{eff}} = 1/2$ are observed, as denoted on the top. **b,** The FQH states local gaps extracted in the $N = 2$ LL. $(\Delta_e + \Delta_h)/2$ (blue and orange bars) and the lower bound of the thermodynamic gap $eS_G \delta V_G = e(S_{G,e} \delta V_{G,e} + S_{G,h} \delta V_{G,h})/2$ (purple squares) are defined following same schematic shown in Figure 3 and Figure 4. Error bars are added to represent the measurement's resolution.

**Methods**

Sample fabrication:

The BLG devices are fabricated with a modified dry transfer method. A polydimethylsiloxane (PDMS) block on a glass slide covered by polyvinyl alcohol (PVA) coated transparent tape is used as a pickup handle. The flakes used are mechanically exfoliated and picked up by the handle sequentially, following the order of bilayer graphene, hBN dielectric (40-60 nm), then a few-layer graphite gate (5-10 nm) lastly. The stack is then deposited to a pre-patterned $SiO_2$(285 nm)/Si substrate with Au/Ti contacts. The PVA film is washed off by using HPLC water, acetone, isopropyl alcohol (IPA), and n-methyl-2-pyrrolidone (NMP) until its residues are completely cleaned off. The device is then transferred into the ultra-high vacuum (UHV) chamber to anneal at 475 °C overnight before putting into STM *in situ*.

STM/STS measurements:

STM/STS experiments are performed in a home-built dilution refrigerator STM with mixing chamber temperature around 20 mK and effective electron temperature around 210 mK calibrated on Al(100). Data in this paper are obtained with a perpendicular magnetic field ranging from 0 T to 13.95 T. The measurements are performed with a tungsten tip prepared on Cu(111) surface, as previously reported in (*21, 60*). In our experiments, the STM tip is grounded and the bias voltage $V_B$ is applied to the BLG. A total of $V_B + V_G$ is applied to the graphite back gate to maintain a relative $V_G$ difference between the sample and the back gate.

For the spectroscopic measurements, the differential conductance d$I$/d$V$ is taken by a lock-in method with a.c. modulation at a frequency of 712.9 Hz. A typical set point of $V_B$ = 0.4 V, $I$ = 2 nA, with a.c. modulation 1 mV is applied for measuring the quantized Landau levels in Fig. 1d in the main text and in Fig. S4. A typical set point of $V_B$ = -0.2 V, $I$ = 2 nA, with a.c. modulation 0.2 mV is applied for measuring the fractional quantum Hall (FQH) states.

For the atomic scale wave function imaging experiments, we collect the data at a constant tip-sample distance. The plane tilt of the measurements area is carefully adjusted, in order to ensure measurements on a flat trajectory with a constant tip-sample distance. We first turned off the feedback loop at a reference initial setpoint (both $V_G$ and $V_B$), and change $V_G$ to the target value, and then lower the bias setpoint $V_B$ to the target value and start the scan with feedback loop turned off. After the scan, we sweep the $V_G$ and $V_B$ back to the reference setpoint before turning the feedback loop on. Measurements are performed after sufficient waiting time until no atomic drift is observed. The spatial dependence of tunneling current $I(V_{excit})$ shows where the wavefunctions locate on the bilayer graphene lattice. Here an integrated total current $I(V_{excit}))$ is used to present the ground state.

Estimation of the tip-gating effect:

Despite the work function mismatch between the STM tip and the BLG surface can be significantly reduced by preparing the tip on the Cu(111) surface, a relative potential difference from the tip bias $V_B$ will still make the tip locally gate the BLG sample. Therefore, an estimation of how much the tip gate the sample will be essential for determining the carrier density in the BLG. The charge doped by the tip is $Q_t = C_{\text{tip-sample}} \cdot V_B$, where $C_{\text{tip-sample}}$ is the capacitance between the STM tip and the sample. We can obtain $C_{\text{tip-sample}}$ by investigating gate-tunable scanning tunneling spectroscopy with a large $V_B - V_G$ plane: when the system first enters a single particle gap ($\nu = \pm 4, \pm 8$, etc.), the transition density from a compressible state to an incompressible one will follow a constant line in the $V_B - V_G$ plane, at higher $V_B$ higher $V_G$ is needed to enter the incompressible state (black line in Fig. S24). The slope of the black line follows the relationship: $1/slope = \frac{\delta V_G}{\delta V_B} = \frac{C_{\text{tip-sample}}}{C_{\text{gate-sample}}} = \frac{7.5}{76} \approx 0.098$. Therefore, while the STM tip is only a few angstroms away from the sample in the tunneling regime, its effective $C_{\text{tip-sample}}$ is very small, showing a minimum tip gating effect. The carrier density (or filling factor) of the BLG sample is dominantly determined by the back gate voltage $V_G$.

We then discuss the potential effect of tip-gating on the size of the FQH gaps. At fractional fillings, the system is more conducting. Thus, the screening from electrons will further reduce the tip-gating effect to render it less than 10%. Another evidence that our tip-gating is less than 10% in the FQH state is all the FQH state features appear at the densities solely controlled by the back gate. Additionally, in a recent study we have even shown that we can image the Wigner crystal with the correct periodicity determined from the carrier density tuned by the back gate voltage by STM in the same samples as this work (*61*), again indicating the tip-gating effect is negligible in our experiment. Nonetheless, we use the 10% tip-gating amount as an upper bound for the tip-gating effect in the FQH states and see how much will this change our results on $\Delta_e$, $\Delta_h$, $\Delta_t$, $\delta V_{G,e}$, $\delta V_{G,h}$, and the thermodynamic gap $\delta\mu = eS_G\delta V_G$. We note that the tip-gating effect could only skew the $V_G$-axis, which will leave the gap size based on the $V_B$-axis unchanged ($\Delta_e$, $\Delta_h$, $\Delta_t$). And the $\delta V_G$ gaps will be changed under

$$\delta V'_{G,e/h} = (C_{\text{gate-sample}} \cdot \delta V_{G,e/h} + C_{\text{tip-sample}} \cdot \Delta_{e/h})/(C_{\text{tip-sample}} + C_{\text{gate-sample}})$$

And since the resulting correction of the lower bound of the thermodynamic gap $\delta\mu' = eS_G'\delta V_G'$ translates into the $V_B$ axis, we expect it to be minimally affected by including the tip gating effect, see Table S13 for the extraction in $0 < \nu < 1$ after considering 10% tip-gating effect. As shown in Table S13, the change of the lower bound of the thermodynamic gap $\delta\mu = eS_G\delta V_G$ after considering tip-gating is below 2.5% for all the FQH states. We therefore are confident that the maximal effect of tip-gating, determined from the ratio of capacitance measured near the IQH state in our experiment (since we expect near FQH states the screening is stronger due to the quantum capacitance), does not impact the conclusion of our paper.

Filling factor assignment

The filling factors of the FQH states are assigned by examining the spectroscopic feature within each LL sector to minimize quantum capacitance effects. A linear fit between the bottom gate voltage $V_G$ and the filling factor $\nu$ is obtained by anchoring the gap features corresponding to the most pronounced FQH states like $\nu = $ 1/3, 1/2, 2/3, etc. This linear fit is then used to determine the exact filling factor of the weaker gap features up to a precision of $\Delta \nu = \pm 0.002$. Extra caution is taken while determining FQH states up to 7/15, 8/17, 9/17, and 8/15, where the linear fit is obtained by considering their nearby identified FQH states to further avoid deviation caused by the quantum capacitance effect inside the gaps.

Testing the validity of the FQH state features in the tunneling spectra

A natural question that arises is whether the observed FQH features are intrinsic to the tunneling conductance itself, or are instead an artifact of the small activated conductivity of the FQH state, which must ultimately carry away the current. As these resistances occur in series, this could result in voltage division of $V_B$ between the tip-sample junction resistance and intrinsic resistance of the sample, thereby influencing FQH features in the STM spectroscopic measurements. To address this concern, we have carried out experiments at different tip-sample separations; due to the exponential dependence of the tunneling resistance on height, this changes tip-sample resistance to that of the sample (See Fig. S9-10). We find all the features of FQH states, and particularly the onset of tunneling current within the incompressible region of gate voltages, to remain independent of the junction resistance (Fig. S9). Unlike transport studies, we are not probing FQH states with low-energy quasi-particles, but with multiple high-energy quasi-particles created by the decay of a tunneling electron (with its energy greater than the Coulomb gap). These quasi-particles are created at energies of several meVs and can transport the small tunneling current to the contacts. The quasi-particle transport also occurs in transport experiments when they are thermally excited and results in finite transport currents at low bias. Since the tunneling currents are small, we are exciting the system with a small density of energetic quasi-particles that decay away from the tip before the next tunneling event.

Extracting the local energy gaps, and the gate width

The tunneling gap $\Delta_t$ of the FQH states is defined as the sum of the energy required to reach a threshold tunneling current at $I_{thre} = 0.5$ pA, denoted as $eV_{B,+}$, for positive bias voltage ($V_B > 0$) and the energy required to reach a threshold tunneling current at $I_{thre} = -0.5$ pA, denoted as $eV_{B,-}$ for negative bias voltage ($V_B < 0$) in the middle of the incompressible gap feature, i.e. $\Delta_t = eV_{B,+} + |eV_{B,-}|$ as defined in the Fig. 3d in the main text. (The $I_{thre} = \pm 0.5$ pA criterion is for $\nu > 0$; for $\nu < 0$ side, $I_{thre}$ is defined to be 0.3 pA and -0.1 pA due to small signal at the top graphene layer in those ranges).

The tunneling Coulomb gap $\Delta_C$ systemically changes with filling factor $\nu$ within a LL sector, as can be identified in Fig. 4a in the main text, as well as in previous reports (*21, 41, 42*). To determine the $\Delta_C$ for extraction of $\Delta_e$ ($\Delta_h$), we determine the linear behavior of $\Delta_C$ with density outside of the incompressible regions in gate voltage and extrapolate the value of $\Delta_C$ for each incompressible gate range for our analysis. Similar to the threshold energy method of the $\Delta_t$ extraction, we define the $\nu_{eff}$ dependent threshold energy $eV_{B,+}(\nu_{eff})$ and $eV_{B,-}(\nu_{eff})$ and linearly fit the values within a proper $\nu$ range to minimize disturbance from FQH states ($0.48 < \nu_{eff} < 0.52$ in the $N = 0$ LLs, $0.35 < \nu_{eff} < 0.65$ in the $N = 1$ LLs, and $0.47 < \nu_{eff} < 0.53$ in the $N = 2$ LL). By this criterion, we are able to determine the tunneling Coulomb gap $\Delta_C(\nu)$ at all the non-integer fillings.

Here we justify our choice of linearly fitting the threshold energy, $eV_{B,+}(\nu_{eff})$ and $eV_{B,-}(\nu_{eff})$ determined by the threshold current $I_{thre}$, as a function of $\nu_{eff}$ within a specific range of $\nu_{eff}$ to get the tunneling Coulomb gap $\Delta_C$, and extrapolating the Coulomb gap to a bigger filling factor range. For example, Fig. S25a shows a gate-tunable scanning tunneling spectroscopy in $0.3 < \nu < 0.7$ ($N = 0$) at $T = 210$ mK, with the red lines being the threshold energy determined by the threshold current $I_{thre} = \pm 0.5$ pA. The black line is the fitted Coulomb gap $\Delta_C$ by fitting the red lines within $0.48 < \nu_{eff} < 0.52$ window. It is clear that near some FQH states the red line deviates from the black line, meaning the extrapolation of $\Delta_C$ fails there. We then performed the same measurement at higher temperatures ($T \approx 4$ K) and extract $\Delta_C$ using the same standard, see Fig. S25b. Here the orange line is the threshold energy $eV_{B,+}(\nu_{eff})$ and $eV_{B,-}(\nu_{eff})$, and the black line is the fitted $\Delta_C$. The orange line has an excellent overlap with the black line, in contrast to the case in Fig. S25a. We conclude that the deviation of the threshold energy from the fitted Coulomb gap $\Delta_C$ is due to the interaction effects or the presence of the FQH state. Therefore we suspect the deviation comes from the interaction between the tunneled electron and the fractionally charged quasi-particles (quasi-holes) near the FQH states. This is clearer if we plot the threshold energy for the high temperature and the low temperature together, see Fig. S25c. Therefore, we conclude the linear fitting and extrapolation of the threshold energy within $0.48 < \nu_{\text{eff}} < 0.52$ gives rise to the intrinsic Coulomb gap $\Delta_C$. For $N = 1$ orbital states, because the deviation between the threshold energy and the fitted line is not as large and it's usually linear throughout the filling range of interest, we select a larger $\nu_{\text{eff}}$ range ($0.35 < \nu_{\text{eff}} < 0.65$) as a standard.

For the additional energy required to excite the quasi-particles/quasi-holes inside the FQH states, we define $\Delta_e$ and $\Delta_h$ (see inset of Fig. 3d). For $\Delta_e$, their values are extracted by subtracting the fitted tunneling Coulomb gap $\Delta_{C,+}(\nu_{\text{eff}})$ at the positive bias voltage (+ stands for $V_B > 0$) from the threshold energy $eV_{B,+}(\nu_{\text{eff}})$ upon entering the FQH states (the highest $eV_{B,+}$ within the incompressible region), namely $\Delta_e = eV_{B,+}(\nu_{\text{eff}}) - \Delta_{C,+}(\nu_{\text{eff}})$; for $\Delta_h$, their values are extracted by subtracting the fitted tunneling Coulomb gap $\Delta_{C,-}(\nu_{\text{eff}})$ at the negative bias voltage ($-$ stands for $V_B < 0$) from the threshold energy $eV_{B,-}(\nu_{\text{eff}})$ upon exiting the FQH states (the lowest $eV_{B,-}$ within the incompressible region), namely $\Delta_h = |eV_{B,-}(\nu_{\text{eff}}) - \Delta_{C,-}(\nu_{\text{eff}})|$. See illustration in Fig. 3d. Notice that $\Delta_e$ and $\Delta_h$ appear at different $V_G$.

The gap width $\delta V_{G,e}$, $\delta V_{G,h}$ and the absolute value of the slope for features $S_{G,e}$, $S_{G,h}$ are extracted as detailed below: For $\delta V_{G,e}(V_B > 0)$, we linear fit starting from the gate of the highest threshold energy $eV_{B,+}(\nu_{\text{eff}})$ as increasing the gate voltage until the gate voltage where the threshold energy drops below the tunneling Coulomb gap $\Delta_{C,+}(\nu_{\text{eff}})$ within the FQH gap feature. The linear fit therefore determines the slope $S_{G,e}$ and the gate voltage where it crosses $\Delta_{C,+}(\nu_{\text{eff}})$. The gate voltage difference between the highest threshold energy and the cross point is then defined as the gap width $\delta V_{G,e}$ on the positive $V_B$ side. For $\delta V_{G,h}(V_B < 0)$, we linear fit starting from the gate of the lowest threshold energy $eV_{B,-}(\nu_{\text{eff}})$ as decreasing the gate voltage until the gate voltage where the threshold energy raises above the tunneling Coulomb gap $\Delta_{C,-}(\nu_{\text{eff}})$ within the FQH gap feature. The linear fit therefore determines the slope $S_{G,h}$ and the gate voltage where it crosses $\Delta_{C,-}(\nu_{\text{eff}})$. The gate voltage difference between the highest threshold energy and the cross point is then defined as the gap width $\delta V_{G,h}$ on the negative $V_B$ side.

We extract the thermodynamic gap $\delta\mu$ from $\delta V_{G,e}$, $\delta V_{G,h}$, $S_{G,e}$, $S_{G,h}$ following the process below. The local measurements of $(\delta V_{G,e} + \delta V_{G,h})/2$ can be analyzed to characterize $\delta\mu/e$ and the "thermodynamic gap" for various FQH states even in the realistic setting where there is a background of impurities in the sample. However, some clarifying remarks are first in order, as the meaning of a "thermodynamic gap" is subtle in the presence of disorder. The chemical potential is the internal free energy required to add charge $-e$ *anywhere* in the sample, and is thus a spatially average quantity as measured in *e.g.* bulk capacitance measurements (*28*, *40*). In FQH states charge enters in the form of fractionalized quasi-particles, and in a perfectly clean system, we thus expect a jump of $\delta\mu = (e/e^*)\Delta_{qp}$ across the FQH state, where $\Delta_{qp}$ is the energy to nucleate a quasi-particle / quasi-hole pair of fractional charge $e^*$. In the presence of disorder, the FQH state acquires a small but finite compressibility due to in-gap impurity states, and thus strictly speaking, the (global) thermodynamic gap is zero. Nevertheless, due to the low impurity concentration, $\mu(n)$ still exhibits a rapid jump across the FQH state, flanked by regions of negative compressibility, and we may operationally *define* $\delta\mu$ by this jump. As discussed in (*52*), this $\delta\mu$ is somewhat reduced from the thermodynamic gap of the clean system and depends on the disorder distribution. In STM however, we only probe the state in the vicinity of the tip, and we have explicitly selected a region in which no impurities are observable within at least 100 nm of the tip (see SI section V and Fig. S16). We first contend that $\delta V_G$ corresponds to the gate range in which the quasiparticle occupation remains unchanged within $\sim \ell_B$ distance from the tip. This interpretation is implied by the fact that within this range, the spectroscopic features simply shift linearly in energy with $V_G$, consistent with the only change being a shift in the potential $\phi$ in the vicinity of the tip. The jumps $\Delta_{e/h}$ then occur when the quasi-hole/particle occupation changes under the tip, changing the energetics of the initial and final tunneling states. We thus expect $\delta V_G$ corresponds to the condition $e^*\delta\phi \approx \Delta_{qp}$, where $\Delta_{qp}$ is the energy required to nucleate a quasiparticle -quasihole pair of fractional charge $e^*$ in the clean system. In the clean limit, $\delta\phi \sim \delta V_G$, and we thus obtain $e\delta V_G = (e/e^*)\Delta_{qp}$. However, with disorder $\delta\phi \neq \delta V_G$, because the local potential is sensitive to the filling of even rather distant impurity states (the Coulomb potential falling off as $d_G^2/r^3$ in our single-gate sample, where $d_G$ is the distance to the graphite gate). Instead, we expect $\delta\phi = S_G \delta V_G$ for some $S_{G,e/h} < 1$ that depends on the impurity density. We interpret the observed slope $S_{G,e/h}$ (around 0.6 for the $\nu = 2/3$ state, for example) of the in-gap spectroscopic features as quantifying precisely this effect, as it relates the reduction in energy for a charge-$e$ excitation ($\phi$) to the applied $V_G$. Within this interpretation $S_G \delta V_G$ becomes a proxy for $\delta\mu$ in the clean limit, though it may be reduced somewhat due to the screening of the interaction by the tip and other effects, the analysis of which we leave to future work.

In practice, we find that $S_G$ and $\delta V_G$ as measured on the e/h side are different so we report the average $eS_G \delta V_G = e(S_{G,e}\delta V_{G,e} + S_{G,h}\delta V_{G,h})/2$ in the figures as the lower bound of the thermodynamic gap $\delta\mu$.

In the extraction of the local energy gaps, we notice sometimes a double or even multiple gap feature for a single FQH state. This can be seen, for example, as the gap feature for the $\nu = 1/3$ FQH state in Fig. 3a in the main text. These gap features could be signatures of in-gap states being filled at remote disorders outside of our measurement field of view. These features affect a precise extraction of the local gaps, for example, causing underestimation of $\Delta_t$ and overestimation of $\delta V_G$.

An example for extracted values $\Delta_t, \Delta_e, \Delta_h, \delta V_{G,e}, \delta V_{G,h}, \Delta_{C,+}, \Delta_{C,-}$ overlay on $\nu = \frac{3}{2}$ state is shown in Fig. S26.

**Data Availability:** Source data for the main figures are provided. Other data that supports the findings of this study are available from the corresponding author upon request.

**Code Availability:** The code that supports the findings of this study is available from the corresponding author upon reasonable request.

**Method-only references:**

**Additional information:**

Supplementary Text

Figs. S1 to S27

Tables S1 to S13